\documentclass[a4paper,11pt]{article}
\usepackage[english]{babel}
\usepackage[T1]{fontenc}
\usepackage{amsmath}
\usepackage{inputenc}
\usepackage{amsthm}
\usepackage{amssymb}
\usepackage{setspace}
\usepackage{enumerate}
\usepackage{subfloat}
\usepackage[eulergreek]{sansmath}
\usepackage{lineno}
\usepackage{soul,xcolor}
\usepackage{xcolor}
\usepackage{array}
\usepackage{microtype}
\usepackage[flushleft, para]{threeparttable}
\usepackage{mathrsfs}
\usepackage{eucal}
\usepackage{bbm}
\usepackage{multirow}
\usepackage{hyperref}
\usepackage{listings}
\usepackage{longtable}
\usepackage{dsfont}
\usepackage{graphicx}
\usepackage{rotating}
\usepackage{tabularx}
\usepackage[percent]{overpic}
\usepackage{varioref}
\usepackage[blocks]{authblk}
\usepackage[font=footnotesize,format=hang]{caption}
\usepackage[a4paper,top=3cm,bottom=3cm,left=2.5cm,right=2.5cm,%
bindingoffset=5mm]{geometry}
\theoremstyle{remark}

\newcolumntype{M}[1]{>{\centering\arraybackslash}m{#1}}
\newcolumntype{L}[1]{>{\arraybackslash}m{#1}}
\newcolumntype{N}{@{}m{0pt}@{}}

\usepackage{amssymb,amsmath,amsfonts,latexsym} 
\usepackage{hyperref}

\usepackage{mathtools}
\usepackage{graphicx}
\usepackage{epstopdf}
\usepackage{float} 
\usepackage{color}
\usepackage{graphics}
\usepackage{booktabs}
\usepackage{bbm}
\usepackage{url}

\theoremstyle{theorem}
\newtheorem{theorem}{Theorem}

\begin{document}

\title{\textbf{A fractional approach to study the pure-temporal Epidemic Type Aftershock Sequence (ETAS) process for earthquakes modeling}}

\author[$^1$]{Lorenzo Cristofaro}
\author[$^2$]{Roberto Garra}
\author[$^3\;\star$]{Enrico Scalas}
\author[$^4$]{Ilaria Spassiani}

\affil[$\star$]{Corresponding author: e.scalas@sussex.ac.uk, enrico.scalas@uniroma1.it}
\affil[$^1$]{Dipartimento di Scienze Statistiche, Universit\`a di Roma ``La Sapienza'', Rome, Italy}
\affil[$^2$]{Section of Mathematics, Universit\`a Telematica Internazionale Uninettuno, Rome, Italy }
\affil[$^3$]{Department of Mathematics, University of Sussex, Falmer, Brighton, UK and Dipartimento di Scienze Statistiche, Universit\`a di Roma ``La Sapienza'', Rome, Italy}
\affil[$^4$]{Istituto Nazionale di Geofisica e Vulcanologia (INGV), Rome, Italy}

\date{\vspace{-5ex}}

\maketitle

\begin{abstract}
In statistical seismology, the Epidemic Type Aftershocks Sequence (ETAS) model is a branching process used world-wide to forecast earthquake intensity rates and reproduce many statistical features observed in seismicity catalogs. In this paper, we describe a fractional differential equation that governs the earthquake intensity rate of the pure temporal ETAS model by using the Caputo fractional derivative and we solve it analytically. We highlight that the tools and special functions of fractional calculus simplify the classical methods employed to obtain the intensity rate and let us describe the change of solution decay for large times. We also apply and discuss the theoretical results to the Japanese catalog in the period 1965-2003. 

\end{abstract} 

\section{Introduction}
\label{sec:Introduction}
The $20^{\mathrm{th}}$ century saw a great expansion for the theory of fractional calculus, which indeed found increasing applicability in many scientific fields, such as bioengineering, physics and rheology (\emph{e.g.} \cite{carcione,baleanu:2012,hilfer:2000,magin:2010,mainardi:2010}). Some recent papers have been devoted to the applications of fractional differential equations in modelling the temporal decay of aftershocks, we refer for example to \cite{sanchez:2019,kong:2021}. De facto, fractional derivatives seem to universally appear in mathematical models of epidemic processes (\emph{e.g.}, see \cite{ascione,balzotti:2020,balzotti:2021,monteiro:2021}), thus playing an important role in handling diffusion and memory mechanisms. In particular, the Caputo fractional derivative is a very useful tool to describe natural processes with memory and an underlying power-law behavior, as it is defined by the convolution between a power law kernel and the ordinary derivative of a function. Therefore, it represents a natural candidate to handle epidemic-type models in seismology, and specifically their decreasing power-law modeling typically used for aftershocks decay. This is indeed the case of a benchmark model in this field: the well-known Epidemic Type Aftershocks Sequence (ETAS) model  \cite{ogata:primo,ogata:secondo,ogata:terzo,ogata:quarto}, belonging to the class of self-exciting, branching, Hawkes processes. In that model it is assumed that any seismic event may generate its own offsprings independently of any other shock in a cascade process. According to the pure temporal ETAS model, the aftershocks rate $\phi_{m_0}\, dt$ generated by a given initial \emph{background} event, occurred at $t=0$ and with magnitude $m_0$, is 

\begin{align}
\label{eqn:omoriutsu}
\phi_{m_0}(t)dt:=&\;C_{m_0}\,\phi(t)\,dt,\quad\text{where $t>0$ and}\\
\label{eqn:omoriutsu000}
C_{m_0}=&\;K_0\,e^{\alpha(m_0-m_{tr})},\\
\label{eqn:omoriutsu111}
\phi(t)=&\;\theta\, t_0^{\theta}\,\frac{1}{t^{1+\theta}}H(t-t_0).
\end{align}
 
We stress that in the ETAS model the events are of two types: \emph{background} if they have not been triggered by any previous shock, and \emph{aftershocks} if they have been generated by a ``mother'' event. In the above expressions~\eqref{eqn:omoriutsu}-\eqref{eqn:omoriutsu000}-\eqref{eqn:omoriutsu111}, $H(\cdot)$ is the Heaviside function and $m_{tr}$ is the completeness magnitude, that is the threshold such that all the events with a higher magnitude are surely recorded in the earthquake catalog. The time-magnitude separable function $C_{m_0}\phi(t)$ is instead the so called Omori-Utsu model describing the decay of aftershocks in time, also known as the \emph{modified Omori law} (MOL, \cite{omori:1894,utsu:1957,otsuka:1987}). Its magnitude component $C_{m_0}$ is named \emph{productivity law}. The parameter $t_0$ finally describes the lapse-time (dead time) immediately after the main shock at 0, within which this event cannot produce its aftershocks and the Omori relation cannot be applied, so $t>t_0$ in~\eqref{eqn:omoriutsu111}. As the intuition suggests, $t_0$ is very small \cite{utsu:1995}. 

In this paper we follow a fractional approach to explicitly derive and analytically solve the self-consistency equation of the pure-temporal ETAS model. This methodology represents quite a novelty in statistical seismology and, differently from the classical approaches based on the Laplace transform \cite{sornettesornette:1999,helmstetter:2003,spassiani:2016}, it allows to obtain a new closed representation of the rate function through the special functions of fractional calculus. Indeed, in this work we illustrate the usefulness of fractional calculus tools to treat problems in statistical seismology. 

For the sake of simplicity, we shall focus here on a single aftershock sequence, triggered by the fixed background event $(t_0=0,m_0)$, which is assumed to generate its offsprings according to the pure-temporal ETAS rate~\eqref{eqn:omoriutsu}-\eqref{eqn:omoriutsu000}-\eqref{eqn:omoriutsu111}. The explicit solution we shall find for the self-consistency equation will also be analyzed asymptotically to investigate any change of regime in the aftershocks decay, in relation to the temporal scale considered (short/long term with respect to the background event) and to the reciprocal order relationship between the magnitudes of mother and daughters.

In what follows, we will first go into the detail of the MOL law for aftershocks (see Section~\ref{sec:omori}), since it acts as a go-between to introduce fractional calculus in seismic modeling analysis. It is also the specific object to investigate for possible changes of regime. In Section~\ref{sec:scon}, we then proceed with the explicit derivation of the ``single'' self-consistency equation of the pure-temporal ETAS, that is, the one relative to the single background-aftershocks sequence we are focusing on. The analytical solution in terms of the rate is therefore obtained in Section~\ref{sec:sol} by means of results of fractional calculus, and asymptotically analyzed in Section~\ref{sec:asy}. A practical application to a real earthquake catalog is then illustrated in Section~\ref{sec:appl}, where we also included some comments about the ongoing seismic sequence behavior in relation to the magnitudes involved. We eventually discuss the results obtained.



\section{The Omori-Utsu model for aftershocks}
\label{sec:omori}
Omori \cite{omori:1894} originally proposed a decreasing power law to model the aftershocks decay in the case of the 1891 Nobi earthquake of magnitude 8. In its first formulation, the law was purely temporal with exponent 1, while the magnitude term $C_{m_0}$ was absent. The introduction of the exponent parameter $\theta+1$ and the inclusion of the magnitude term are due to Utsu \cite{utsu:1957} and Otsuka \cite{otsuka:1987}, respectively. We stress that $C_{m_0}$ controls the average number of aftershocks generated by each event, that is, the so-called \emph{branching ratio}. 

MOL is actually the most used model in practical applications, but, in the last decades, a debate has ignited in the literature about the possibility that a single MOL, generated by an initial backgorund event, may or may not evolve into a global modified Omori law, that models the entire seimic process developing in successive generations. The factors that may indeed induce a break in the sequence's decreasing trend, triggered by the initial shock, are the potential occurrence of aftershocks with a magnitude comparable to the one of the backround event, and/or the passage of a sufficiently long time since the latter's occurrence. To account for these considerations, since 1894, many models alternative to MOL have been proposed, mainly of power law, exponential, stretched exponential and Gamma types (e.g., see \cite{mogi:primo,burridge:primo,souriau:primo,ogata:primo,dieterich:primo,gross:primo,sornhelmn:primo,lolli:primo}). For example, in Mignan \cite{mignan:primo} and \cite{mignan:secondo}, the author finds that, when using the complementary cumulative distribution function of the MOL (i.e. when using the rank plot representation), the stretched exponential model gives a better fit to the whole seismic aftershocks decay. In particular, he obtains this behavior for the 1891 Great Nobi earthquake, that is exactly the dataset used in the landmark article by Omori \cite{omori:1894}. Mignan's argument has been considered misleading by Hainzl and Christophersen \cite{hainzl:primo}; these authors criticize the comparison of finite data sets with functions integrated over infinite periods, and find that MOL is preferable when considering the Maximum Likelihood Estimates (MLEs). Nevertheless, the ambiguity of the MLEs noted by Mignan \cite{mignan:terzo} due to the specific structure of the MOL suggests that one has to consider both statistical results and physical features to choose the right temporal aftershock decay model.

Results by Mignan \cite{mignan:primo} and Hainzl and Christophersen \cite{hainzl:primo} could be both valid, and the differences may be due to the fact that the problem is analyzed from two different perspectives. In this regard, an interesting point to stress is that Omori \cite{omori:1894} proposed the power law decay to model the global aftershocks sequence of the great 1891 Nobi earthquake; nevertheless, for seismic forecast purposes the modified Omori law is used only for first generation aftershocks following an initial strong event. This is assumed for example in the ETAS model, where the aftershock component of the self-consistency seismic rate is obtained by the superposition of MOLs \cite{ogata:primo,ogata:secondo,ogata:terzo,ogata:quarto}. As the intuition suggests, nothing ensures that a single MOL generated by the starting shock remains valid for the sequence of all the aftershocks, at least in general \cite{spassiani:2018}. 

The papers \cite{sornettesornette:1999} and \cite{sornhelmn:primo} deal with a similar issue in the ETAS context, that is averaging over all the possible earthquake sequences: they study how the aggregate ETAS process matches the modified Omori law. The authors define the regime of criticality through the parameter $\theta$ and the branching ratio $n$, which is obtained as the integral over time and all the magnitudes of the MOL multiplied by the decreasing exponential Gutenberg-Richter law for the events' sizes \cite{gr:1944}. In particular, the authors distinguish three regimes: $n<1$ and $\theta>0$ (subcritical), $n>1$ and $\theta>0$ (supercritical), $\theta<0$ and $n$ infinite. In the subcritical regime they observe a transition from a power law decay with exponent $1-\theta$ to a power law decay with exponent $1+\theta$ (MOL); in the supercritical regime, the authors observe a transition from an Omori power law with exponent $1+\theta$ (MOL) to an explosive exponential increase. In both the two cases above, the transition is marked by the characteristic time $t_{HS}^*$ depending only on the ETAS parameters. In the case of infinite $n$ ($\theta<0$), the authors observe a transition from a power law decay with exponent $1-|\theta|$ to an exponential increase, but it is now marked by a different characteristic time, depending again only on the ETAS parameters.

Interestingly, when discarding the magnitude component in equation~\eqref{eqn:omoriutsu}, that is when focusing on $\phi(t)dt$ normalized by $\bar{\phi}=\int_0^{\infty}\phi(\tau)\,d\tau$, where $\phi(\cdot)$ is the MOL in equation~\eqref{eqn:omoriutsu111}, the self-consistent equation that describes the seismicity rate $N(t)$ in $t$ is the classical Wiener-Hopf integral equation
\[
N(t) = \bar{\phi}\int_0^{t-t_0}N(\tau)\theta t_0^{\theta}\frac{1}{(t-\tau)^{1+\theta}}d\tau,
\]
where we recall that the seismicity rate is ``the number of earthquakes in a specified interval of space-time-magnitude, normalized by the length of the time interval'' (see \cite{corssa}). Under certain conditions, it can be viewed also as a fractional integral equation similar to that of a fractional linear death model \cite{sornettesornette:1999,garra:2011}. This suggests the use of the theory of fractional calculus to deal with the ``single'' self-consistency equation of the pure-temporal ETAS model, and taking advantage of the straightforward character of the results to investigate the range of validity of MOL in different conditions. 

\section{Fractional approach to deal with the ``single'' self-consistency equation of the pure temporal ETAS model}
\label{sec:scon}

As promised in the introduction, we now derive the ``single'' self-consistency equation for the pure temporal ETAS model.

\subsection{From the conditional intensity to the mean} 
We can define the fractional ETAS model by its intensity, based on the history of occurences $\mathcal{H}_t=\{t_i\in [0,+\infty) | t_i<t \}$:

\begin{align}\label{defint}
\lambda(t|\mathcal{H}_t)&=\phi_{m_0}(t)+\int_{0}^{t}\phi_{m_\tau}(t-\tau)\mathcal{N}(d\tau) \\
&= \phi_{m_0}(t)+\sum_{t_i<t}\phi_{m_\tau}(t-t_i).
\end{align}
where $\phi_{m_0}$ and $\phi_{m_\tau}$ are defined in (\ref{eqn:omoriutsu}) with $m_0$ the magnitude of the first earthquake and $m_\tau$ the magnitude of the earthquake at time $\tau$. As highlighted in the paper by Chen \emph{et al.} \cite{chen:2021}, we stress that the number $\mathcal{N}$ of events in a certain time interval ($dt$) is a branching random variable, with the rate $\lambda(t|\mathcal{H}_t)$ defined above. Note that the $t_i$s in the above equations are random variables. 

\noindent Let us define the expected intensity as
\begin{equation*}
	\label{expint}
	\lambda(t) = \mathbb{E}[\lambda (t|\mathcal{H}_t)].
\end{equation*}
  Now we obtain the self-consistent equation by applying the expectation on (\ref{defint}) leading to:

\begin{equation}
	\label{SCE}
	\lambda(t) = \phi_{m_0} (t) + \int_0^t \phi_{m_\tau} (t-\tau) \lambda(\tau) \, d \tau.
\end{equation}
We can compute $\mathbb{E}[ N(d\tau)]$ using the definition of conditional intensity and conditioning twice, so that $\mathbb{E}(N(d\tau))=\lambda(\tau)d\tau$, and Campbell's theorem to exchange expectation and integral, see \cite{daley:2003}.

\subsection{Solution by Fractional Calculus}
\label{sec:sol}
To solve the self-consistent equation and find the explicit form of $\lambda(t)$ for $t\geq t_0$, we first recall that in our case $N(0)=1$, or equivalently $\lambda(0)=1$, and $\lambda(t)=0$ for $t \in (0,t_0]$. For the sake of simplicity, hereafter we will consider that all the aftershocks have magnitudes equal to the strongest one, say $m_1$, which in fact can be thought as the reference non-negligible contribute to the total rate. Then, setting $m_\tau=m_1$, $B_{m}=K_0e^{\alpha(m-m_{tr})}\theta t_0^\theta$ and $\phi_{m}(t)=B_{m}t^{-1-\theta}H(t-t_0)$ in equation~\eqref{SCE}, we have that: 

\[  \lambda(t)=\phi_{m_0}(t)+B_{m_1}\int_{t_0}^{t}\frac{H(t-\tau-t_0)}{(t-\tau)^{1+\theta}}\lambda(\tau)d\tau. \]

Applying the definition of the Heaviside function, we obtain for $t>t_0$:

\begin{eqnarray*} 
\lambda(t)&=&\phi_{m_0}(t) + \int_{t_0}^{t-t_0}\frac{B_{m_1}}{(t-\tau)^{1+\theta}}\lambda(\tau)d\tau \\
&=&\phi_{m_0}(t)+ \Big[ \frac{B_{m_1}}{\theta}\frac{1}{(t-\tau)^{\theta}}\lambda(\tau) \Big] _{t_0}^{t-t_0}-\frac{B_{m_1}}{\theta}\int_{t_0}^{t-t_0}\frac{1}{(t-\tau)^{\theta}}\lambda'(\tau)d\tau \\
&=& \phi_{m_0}(t) + \frac{B_{m_1}\lambda(t-t_0)}{\theta t_0^\theta}-\frac{B_{m_1}}{\theta}\int_{t_0}^{t-t_0}\frac{1}{(t-\tau)^\theta}\lambda'(\tau)d\tau. 
\end{eqnarray*}

Recalling the meaning of the parameter $t_0$, we assume now that $t_0 \ll 1$ and we look for continuous solutions so that $\lambda(t -t_0) \approx \lambda(t)$. Furthermore, since we expect that, for large $t$, $\lambda(t)\approx \lambda(s)$ for $s\in[t-t_0,t]$, we can consider $\lambda'(s)\approx0$ for $s\in[t-t_0,t]$. In addition to this, we have $\int_{t-t_0}^t \frac{1}{(t-\tau)^\theta}d\tau=\frac{t_0^{1-\theta}}{1-\theta}$ and $\lambda(t_0)=0$. We finally stress that, since $t_0 \ll 1$, the boundedness of the integrand guarantees that the integral can be extended to $t$ without a large error.

These considerations allow us to assume that $\int_{t-t_0}^t\frac{1}{(t-\tau)^\theta}\lambda'(\tau)d\tau \sim 0$ for large values of $t$, leading to:
\[  \lambda(t)=\phi_{m_0}(t)+\frac{B_{m_1}\lambda(t)}{\theta t_0^\theta} - \frac{\Gamma(1-\theta)B_{m_1}}{\theta}(^cD_{t_0+}^\theta \lambda)(t), \quad t>t_0.\]
Hence, from \eqref{SCE} using the approximations described above, we obtain the fractional differential equation which describes the expected intensity $\lambda(t)$ for $t\geq t_0$: 

\begin{equation}
\label{C}
	\begin{cases*}
		 (^cD^\theta_{t_0+}\lambda )(t) -\nu \lambda(t)=\frac{\theta}{B_{m_1}\Gamma(1- \theta)}\phi_{m_0}(t)\qquad \text{ for } t > t_0,\\
		 \lambda(t)=0 \qquad \qquad \qquad \qquad \qquad \qquad \qquad \qquad \text{for } t=t_0,
	\end{cases*}
\end{equation}
where $\nu=\frac{\theta}{B_{m_1}\Gamma(1-\theta)}(\frac{B_{m_1}}{\theta t_0^\theta}-1) $ and $^cD^\theta_{t_0+}$ is the Caputo fractional derivative. \\
For $a>0$, $\alpha \in (0,1)$ and a suitable function $\psi$ we recall that the Caputo fractional derivative is defined as follows:

\begin{equation*}  (^cD^\alpha_{a}\psi ) (x)=\frac{1}{\Gamma(1-\alpha)}\int_{a}^x \frac{\psi'(s)}{(x-s)^{\alpha}}ds.  
\end{equation*}

 The solution of the fractional differential equation \eqref{C} is given by Theorem \ref{KilThm}, in Appendix 1 (see \cite{kilbas:primo}):

\begin{equation}
\label{eqn:solution}  
\lambda(t)=\frac{\theta B_{m_0}}{B_{m_1}\Gamma(1- \theta)}\int_{t_0}^t (t-\tau)^{\theta-1}E_{\theta,\theta}(\nu(t-\tau)^{\theta})\tau^{-1-\theta}d\tau , \quad t \geq t_0
\end{equation}
where:

\[ 
B_{m_1}=K_0e^{\alpha(m_1-m_{tr})}\theta t_0^\theta; 
\]

\begin{equation}
\label{eqn:nu}
\nu=\frac{\theta}{B_{m_1}\Gamma(1-\theta)}\left( \frac{B_{m_1}}{\theta t_0^\theta}-1\right)=\frac{K_0 e^{\alpha(m_1-m_{tr})}-1}{K_0 t_0^\theta \Gamma(1-\theta)e^{\alpha(m_1-m_{tr})}}
\end{equation}
 and 
\[   
 E_{\theta,\theta}(\nu s^{\theta})= \sum _{k\geq0} \frac{(\nu s^{\theta})^k}{\Gamma(k\theta+\theta)}
\]
is the two-parameter Mittag-Leffler function, see \cite{gorenflo:2014}. We notice that $\lambda(t) \to 0$ for $t \to t_0^+$. Besides, if the constant $\nu$ is negative, the Mittag-Leffler function with two parameters and the solution $\lambda(t)$ decays to zero for $t \to \infty$. In order to have $\nu<0$ we need that $m_1-m_{tr}<-\ln(K_0)\alpha^{-1}$, as $\alpha>0$.

We finally stress that the solution \eqref{eqn:solution} of problem (\ref{C}) is proportional to a convolution, and this fact will be used to determine its asymptotic behaviour. This result will be illustrated in the next section.

\subsection{Asymptotics}
\label{sec:asy}
We focus here on managing the convolution kernel to define the characteristic time $t^*$, which marks the transition of the decaying trend of $\lambda(t)$, see \cite{sornhelmn:primo}. Its behaviour will be compared with the one of the solution found in \cite{sornhelmn:primo}, which is also expressed in the terms of our convolution kernel. 

Let us start by considering the explicit form of the solution (\ref{eqn:solution})
 as a convolution
\begin{equation}
\label{eqn:convol}
\lambda(t)=\frac{\theta B_{m_0}}{B_{m_1}\Gamma(1- \theta)}\cdot\left( t^{\theta-1}E_{\theta,\theta}(\nu(t^{\theta}))\star H(t-t_0)t^{-1-\theta}\right),
\end{equation}
where $ H(t-t_0)$ is the translated Heaviside function.\\

We observe that the functions involved in the convolution in~\eqref{eqn:convol} can be written in terms of two probability densities:
 $-\nu^{-1} t^{\theta-1}E_{\theta,\theta}(\nu(t^{\theta}))$ is the density of a Mittag-Leffler random variable $X_{ML}$, and $\theta t_0^\theta t^{-1-\theta}H(t-t_0)$ is the density of a Pareto random variable $X_P$. Thus, ignoring the constant terms in front of the integral in~\eqref{eqn:solution}, and setting $Z = X_{ML}+X_{P}$ as the sum of the two independent random variables just defined, we can deduce that the solution $\lambda(t)$ of our problem is proportional to the probability density function $f_Z (z)$ :
\begin{equation}
f_Z (z) \propto \int_{t_0}^z (z - t)^{\theta - 1} E_{\theta, \theta} (\nu (z-t)^\theta) t^{-1-\theta} \, d t.
\label{convolution1}
\end{equation}
In order to get the asymptotic behaviour, we first note that we have $\theta \in (0,1)$. It then follows that the Pareto density can be approximated by a Mittag-Leffler density of index $\theta$. As we are interested in the asymptotic behaviour, instead of looking for an explicit approximation, we can just use the fact that the tail of the Mittag-Leffler density vanishes as $t^{-1-\theta}$. Therefore, we replace $X_P$ with another copy of $X_{ML}$ and we study the following integral:
\begin{equation}
f_Z (z) \propto \int_0^z (z - t)^{\theta - 1} E_{\theta, \theta} (\nu (z-t)^\theta) t^{\theta-1} E_{\theta,\theta} (\nu t^\theta) \, d t.
\label{convolution2}
\end{equation}

\begin{figure}[h!]
\centering
\includegraphics[trim=0.01cm 0.01cm 0.01cm 0.01cm,clip,width=\columnwidth]{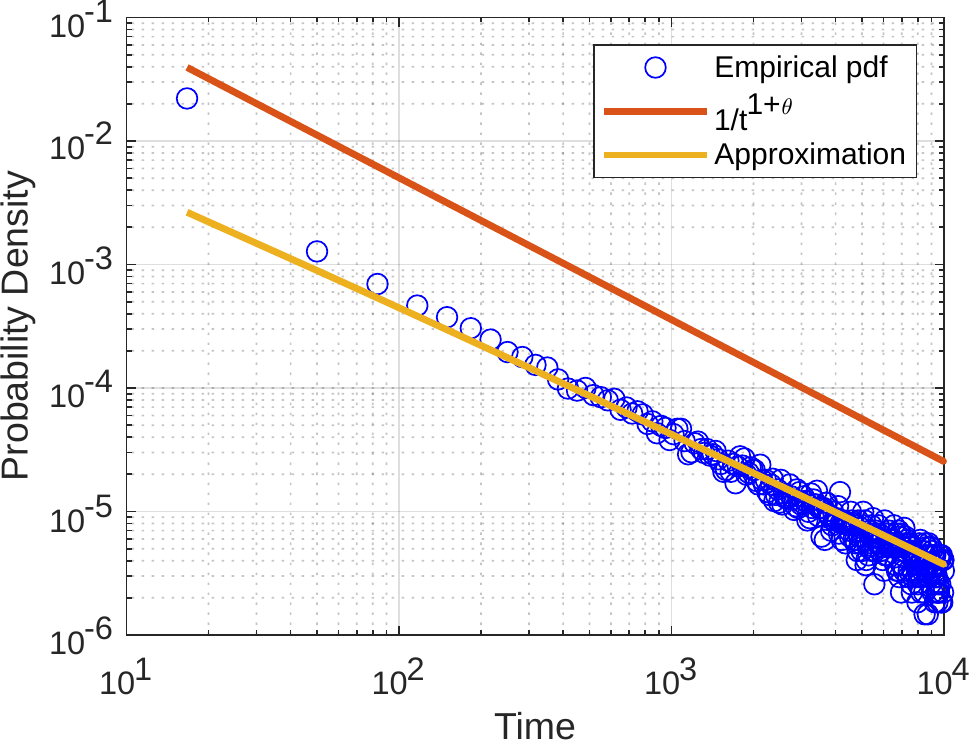}
\caption{Probability density function proportional to the solution $\lambda(t)$. The parameters used are
described in the next section, and they coincide with the parameters taken from the Japanese catalog in the period 1965-2003.}
\label{fig:figure1}
\end{figure}

It can be exactly computed in term of Prabhakar functions (three-parameter Mittag-Leffler functions). They are defined as follows \cite{prabhakar71}
\begin{equation}
E^{\xi}_{\beta,\gamma} (u) = \sum_{r=0}^\infty \frac{(\xi)_r}{r! \Gamma(\beta r + \gamma)} u^r,
\label{prabhakar}
\end{equation}
where the Pochhammer symbol $(\xi)_r$ represents the ascending factorial defined as $(\xi)_r = \xi (\xi +1) \cdots (\xi + r -1)$, with $\xi \neq 0$ and $\beta, \gamma, \xi, u \in \mathbb{C}$ with $\mathrm{Re}(\beta) > 0$. We can now use the following formula \cite{haubold11,cahoy}
\begin{equation}
\int_0^x (x - t)^{\beta -1} E^\gamma_{\alpha,\beta} (a(x-t)^\alpha) t^{\delta -1} E^\sigma_{\alpha, \delta} (at^\alpha) \, dt = x^{\beta+\delta -1} E^{\gamma+\sigma}_{\alpha,\beta+\delta} (a x^\alpha).
\label{convolutionintegral}
\end{equation}
Let us first remark that $E_{\theta, \theta} (u) = E^1_{\theta, \theta} (u)$, then using \eqref{convolutionintegral}, we can conclude that
\begin{equation}
f_Z (z) \propto z^{2 \theta -1} E^2_{\theta,2 \theta}(\nu z^\theta).
\label{result}
\end{equation}
The asymptotic behaviour of Prabhakar functions is studied e.g. in \cite{garra18}. In particular, in our case it turns out that $\gamma = \beta  \xi$ as $\xi=2$, $\beta = \theta$, $\gamma= 2 \theta$. Therefore, the first term in the sum presented in \cite{garra18} cancels out and,
for $t \to \infty$, 
as we have $\nu < 0$, we can see that the dominant behaviour for $t \to \infty$ of $f_Z (z)$ is
\begin{equation}
f_Z (z) \sim z^{2 \theta - 1} z^{- 3\theta} = z^{-1-\theta}.
\label{asymptotic2}
\end{equation}
This gives the asymptotic behavior of our solution $\lambda(t)$ for $t \to \infty$. A graphical representation of our derivation can be found in Fig.~\ref{fig:figure1}, where it is shown that the empirical probability density well retraces the approximation, whose large $t$ trend is the same as that of the function $t^{-\theta-1}$.

\begin{figure}[h!]
\centering
\includegraphics[trim=1.5cm 1.5cm 1.5cm 1.5cm,clip,width=\columnwidth]{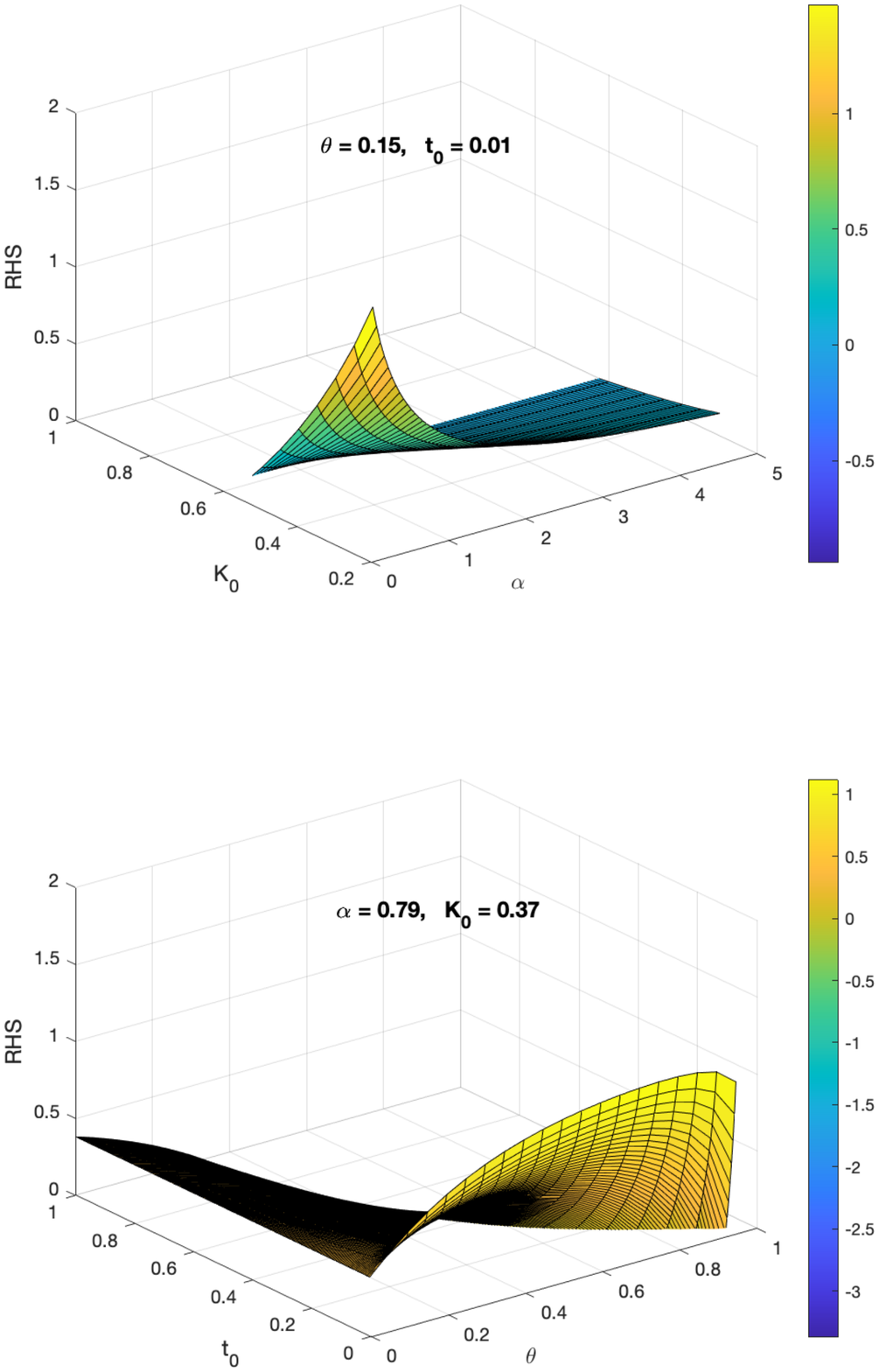}
\caption{3D representation of the right-hand side (RHS) of the last inequality in~\eqref{eqn:rhs} for the characteristic time $t^*$, as a function of $(K_0,\alpha)$, with fixed $(t_0,\theta)$, in the top panel; viceversa, in the bottom one.}
\label{fig:figure2}
\end{figure}

It is also useful to observe that the kernel in the convolution above can be written as 
\[
t^{\theta-1}E_{\theta,\theta}(-\left(t/t^*\right)^{\theta}),
\]
i.e., just rescaling the two-parameter Mittag-Leffler function with respect to the critical time 
\[
t^* = (-1/\nu)^{1/\theta} =t_0\frac{ \left(K_0 \Gamma(1-\theta)e^{\alpha(m_1-m_{tr})}\right)^{1/\theta}}{(1-K_0 e^{\alpha(m_1-m_{tr})})^{1/\theta}},
\]
where we have $\frac{ \left(K_0 \Gamma(1-\theta)e^{\alpha(m_1-m_{tr})}\right)^{1/\theta}}{(1-K_0 e^{\alpha(m_1-m_{tr})})^{1/\theta}} >1$, assuming that $m_1-m_{tr}<-\displaystyle\frac{\ln(K_0)}{\alpha}$. 
Interestingly, $t^*$ only depends on the magnitude $m_1$ and, for any arbitrary constant $\varepsilon$, it holds
\begin{equation}
\label{eqn:rhs}
t^*\le\varepsilon\quad\Leftrightarrow\quad m_1-m_{tr}\le \frac{1}{\alpha}\ln\frac{\displaystyle \left(\frac{\varepsilon}{t_0}\right)^{\theta}}{K_0\left[\Gamma(1-\theta)+\displaystyle \left(\frac{\varepsilon}{t_0}\right)^{\theta}\right]}.
\end{equation}
The right-hand side (RHS) of the last inequality above is plotted in Fig.~\ref{fig:figure2} for $\varepsilon=0.5$ and: as a function of $(K_0,\alpha)$ and fixed $(t_0,\theta)$ in the top panel, viceversa in the bottom one. Since $m_1$ has to be larger than $m_{tr}$, we also plot only positive values of RHS. By looking at the figure, we can deduce that RHS remains small for typical parameters' ranges and, consequently, so does the characteristic time. More precisely, we obtain that $m_1-m_{tr}$ is always smaller than $1.5$ and in this case $t^*<0.5$.

According to the analysis developed in \cite{sornhelmn:primo}, the critical time $t^*$ is particularly relevant, since it describes the cross-over from the ``short time'' Omori's law to the ``long time'' behaviour (see \cite{sornhelmn:primo} for the full discussion). In particular, we have that the behaviour of the kernel $K(t)=t^{\theta-1}E_{\theta, \theta}(\nu t^\theta)$ can be approximated as follows:
 \begin{itemize}
 	\item $K_{t<t^*}(t)\sim 1/t^{1-\theta},$ for $t_0<t<<t^*$,
 	\item $K_{t>t^*}(t)\sim 1/t^{1+\theta},$ for $t>>t^*$.
 \end{itemize}

By looking again at Fig.\ref{fig:figure2}, we can deduce that when $m_1$ is sufficiently small, the kernel can almost always be approximated by the classical MOL (second point above), as $t$ will quite immediately be larger than $t^*$.

\section{Application to a real earthquake catalog}
\label{sec:appl}
\begin{figure}[h!]
\centering
\includegraphics[trim=1.5cm 2cm 1.5cm 0.1cm,clip,width=\columnwidth]{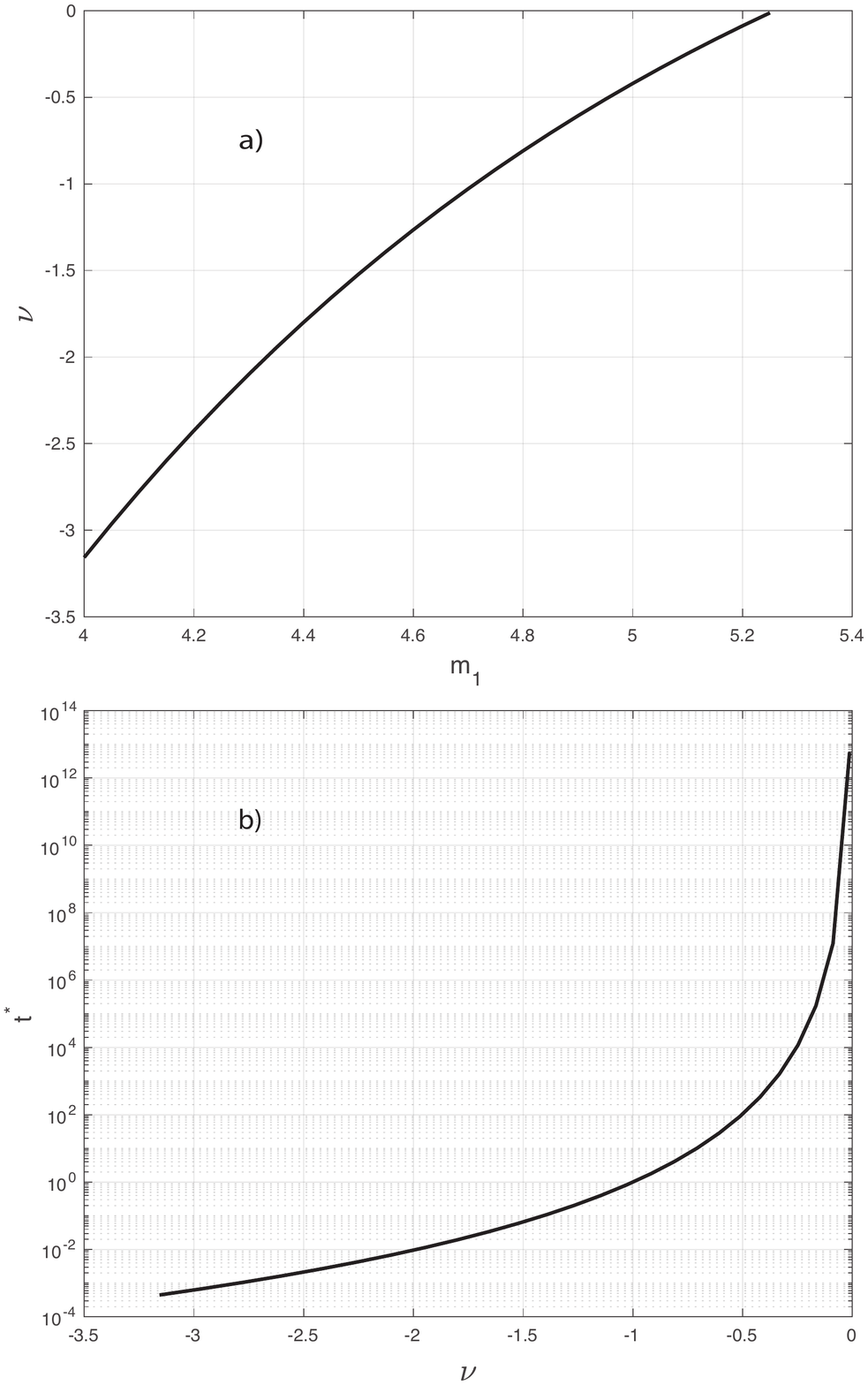}
\caption{Panel a): monotonic behavior of $\nu$ in~\eqref{eqn:nu} as a function of the aftershocks magnitude $m_1$. Panel b): monotonic behavior of the critical time $t^*$ as a function of $\nu$.}
\label{fig:figure3}
\end{figure}

We now turn to explicitly compute the theoretical solution obtained above in a practical case, and to validate the conclusions we discussed. To do that, we consider a set of parameters from the literature, to have a consistent set of estimates, and we fix two arbitrary values for $m_0$ and $m_1$. In particular, we consider the parameters 
$(K_0, \theta, \alpha, t_0) = (0.37, 0.149, 0.79, 0.0078)$ as estimated for the Japanese catalog (JMA) in the period 1965/01/01 - 2003/09/23 by Zhuang \cite{zhuang:2011}. We stress that his $p$ corresponds to our $\theta+1$. The completeness magnitude in this case is $m_{tr} = 4$. With these values for the input parameters, simple computations give that $m^* = 5.26$ is the zero of $\nu$ in~\eqref{eqn:nu}.
Panel a) of Fig.~\ref{fig:figure3} shows that this latter function increases with $m_1$. The same monotonic behavior is observed in panel b) for the critical time $t^*$ as a function of $\nu$. In agreement with the results discussed in the previous Section~\ref{sec:asy}, we conclude that a stronger $m_1$ implies a longer critical time $t^*$, therefore $t<t^*$ for a longer period, and this is the case of deviation from the classical MOL.

In order to guarantee that $\nu<0$ and, consequently, that the solution $\lambda(t)$ does not explode (see Section~\ref{sec:sol}), we have to consider magnitude values smaller than $m^*$. We then fix $m_0 = 4.65$, and we consider the two cases:
\begin{itemize}
\item[$1)$] $4.05 = m_1 < m_0$;
\item[$2)$] $5.25 = m_1 > m_0$.
\end{itemize}
The equality case is instead analyzed for a small and a high value of the magnitudes:
\begin{itemize}
\item[$3)$] $m_0 = m_1 = (4.05;5.25)$.
\end{itemize}

\begin{figure}[h!]
\centering
\includegraphics[trim=0.01cm 11cm 0.1cm 4.5cm,clip,width=\columnwidth]{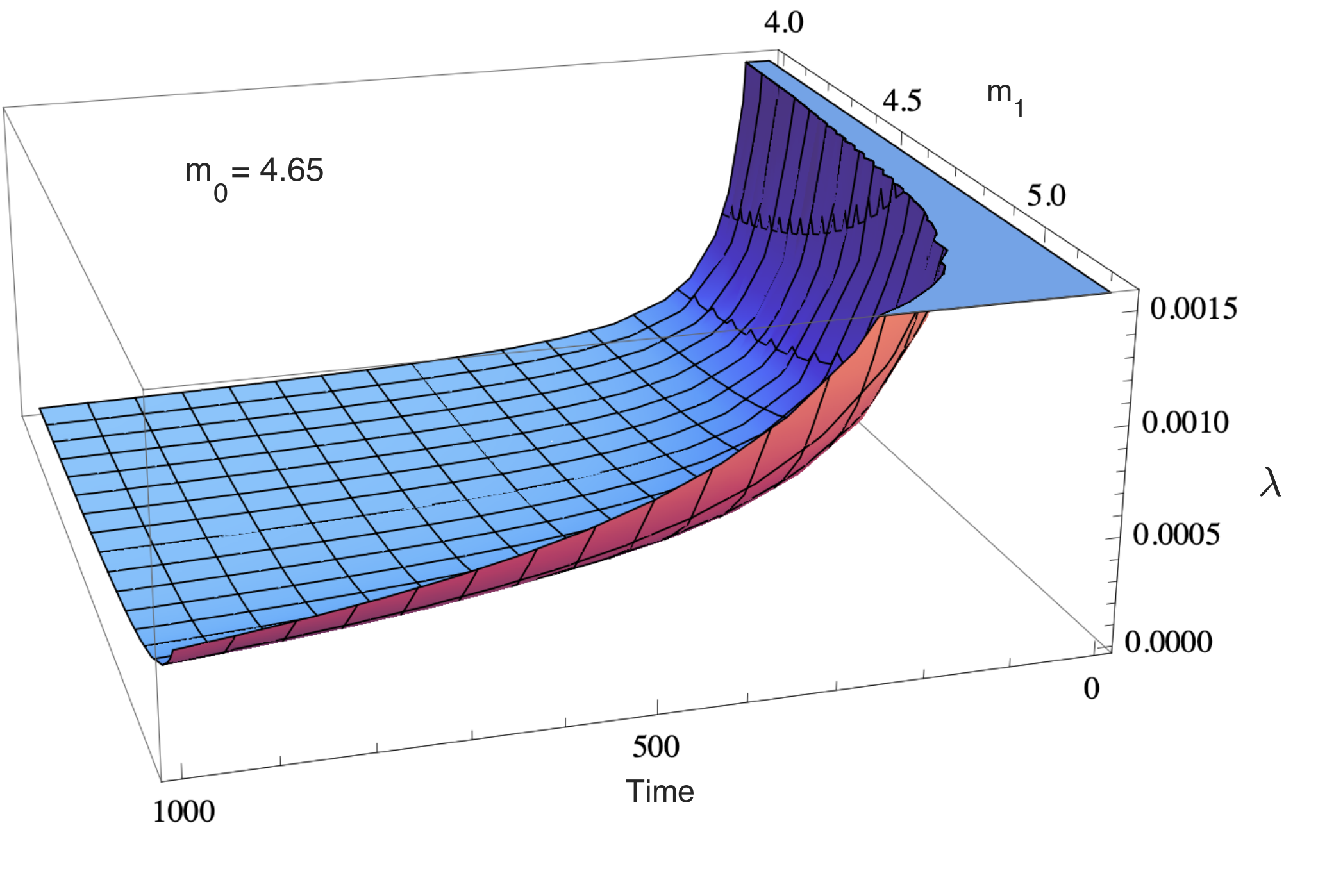}
\caption{3D representation of the global solution $\lambda(\cdot)$, defined in~\eqref{eqn:convol}, as a function of the time $t$ and the aftershocks' magnitude $m_1$. The value of $m_0$ is fixed to 4.65, while the parameters set used is $(K_0, \theta, \alpha, t_0) = (0.37, 0.149, 0.79, 0.0078)$, as estimated in \cite{zhuang:2011}.}
\label{fig:figure4}
\end{figure}

The 3D graphical representation of the solution $\lambda(\cdot)$ in~\eqref{eqn:convol}, as a function of the time $t$ and the aftershocks' magnitude $m_1$, for fixed $m_0 = 4.65$ and the set of parameters by Zhuang \cite{zhuang:2011} introduced above, is given in Fig.~\ref{fig:figure4}. The plot shows a decreasing trend of the solution, that is faster when $m_1<m_0$; this is also the case in which the decrease starts immediately after the dead time $t_0$. For a higher $m_1$, a little more time is instead necessary for the solution to begin its decrease, and this latter has a slightly lower velocity rate. Finally, a sort of plateau appears in correspondence of longer times. This is what expected, as the aftershocks sequence induces an increase in the mean $\lambda(t)=\mathbb{E}(\lambda(t|\mathcal{H}_t))$ proportional to its size, until when, after a sufficiently long time period, it lowers to the background level of seismicity.

\begin{figure}[h!]
\centering
\includegraphics[trim=0.01cm 2.5cm 0.01cm 1.35cm,clip,width=\columnwidth]{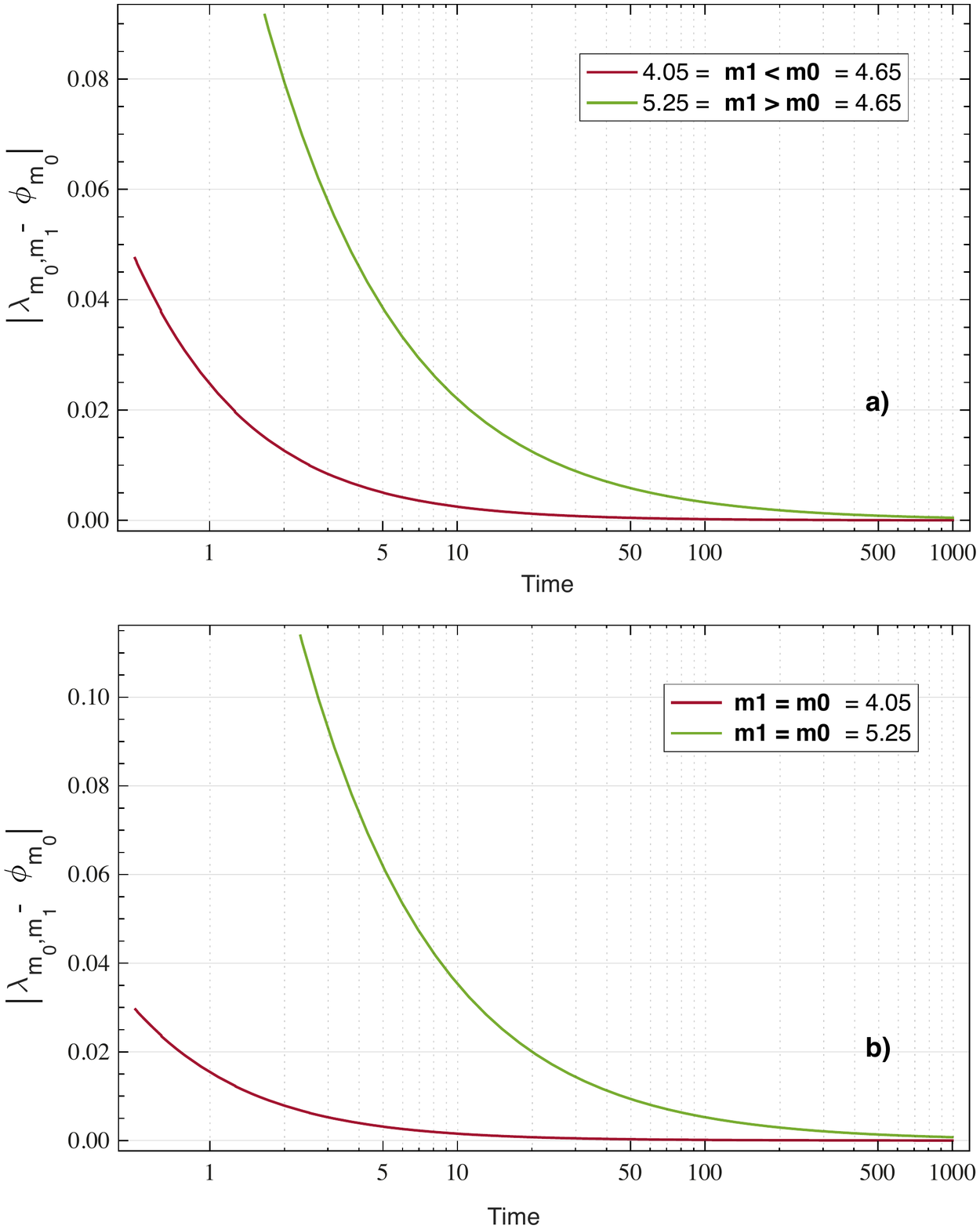}
\caption{Temporal evolution of the absolute difference between the explicit solution $\lambda_{m_0,m_1}(t) := \lambda(t)$, defined in~\eqref{eqn:convol}, and the Omori law $\phi_{m_0}$, for the cases $m_1\gtrless m_0$ and $m_1=m_0$ in panels a) and b), respectively. }
\label{fig:figure5}
\end{figure}

Fig.~\ref{fig:figure5} shows instead the temporal evolution, in $x$-log scale, of the absolute difference between the explicit solution $\lambda_{m_0,m_1}(t) := \lambda(t)$ in~\eqref{eqn:convol} and $\phi_{m_0}$, for the three cases of above: panel $a)$ for the first two, panel $b)$ for the third one. By looking at the plots, we observe that when the aftershock magnitude $m_1$ is strong, even stronger than the $m_0$ of the first event, the global MOL is clearly different than the decreasing power law $\phi_{m_0}$ generated in $t=0$ (indicated in the figure as a dashed line). Instead, the difference is much less evident when $m_1<m_0$. In particular, the mean absolute distance between the global $\lambda_{m_0,m_1}$ and $\phi_{m_0}$ in this latter case is 0.19 within the first 5 days, and 0.17 within the first month. In the same periods, for $m_1 > m_0$, this distance increases to 0.67 and 0.57, respectively. Finally, also the global mean distance increases of more than 200\% when $m_1$ is larger that $m_0$.

The case $m_0=m_1$ is instead linked to the value we select for these magnitudes. If it is low (4.05), we obtain that $\lambda_{m_0,m_1}$ on average deviates from $\phi_{m_0}$ of 0.12 and 0.1 within 5 and 30 days, respectively. These values increase to 1.1 and 0.91 when $m_0=m_1=5.25$. The global mean distance increases instead of one order of magnitude. As expected, this is the case in which we appreciate the largest difference between the global and the initial MOL, and in fact the two high values considered are expected to strongly influence the ongoing seismic sequence.

We stress that the results illustrated above are not conditioned to the specific input setting we selected. In fact, we repeated the same analysis by considering different parameters sets and different geographical regions, still obtaining the same results. 
In the repository \url{https://github.com/FractionalEarthquakes/FractionalEarthquakes} we added the programs we developed to obtain the results presented in this paper.

\section{Discussion and conclusions}
The epidemic-type nature of the ETAS model for seismic sequences, as well as the underlying power-law behavior it considers for the aftershocks' decay, make this model a natural object of study within the fractional theory. In line with this consideration, in this paper we developed a new procedure to derive and solve the self-consistency equation of the pure-temporal ETAS model, in the case of a single earthquake sequence. Differently from the classical approach, based on the Laplace transform and quite laborious to carry through with, our methodology allowed to obtain a closed form of the ETAS rate function in a straightforward way, by means of results of fractional calculus. For the sake of simplicity, we considered here a basic sequence in which a single mainshock, occurred in $t=0$ with magnitude $m_0$, gives birth to its family of aftershocks all with magnitude $m_1$ equal to strongest one. 

Under reasonable approximations in $[t-t_0,t]$ related to the meaning of the dead time $t_0$ (see Section~\ref{sec:sol}), we have shown that the ``single'' self-consistency equation of the pure-temporal ETAS model can be written as a non-homogeneous, differential equation involving Caputo fractional derivative. The explicit solution of such an equation is the result of a well-know theorem of fractional theory \cite{kilbas:primo}, which then allowed us to give the representation of the pure-temporal ETAS rate in a closed form. The function we obtained depends on the two-parameter Mittag-Leffler function, and this agrees with the results found in \cite{sornhelmn:primo}. 

Our next step has been to study the asymptotics of the solution. To do that, we noticed that the solution is proportional to a convolution integral, whose kernel allowed to identify a critical time marking the transition from ``short'' to ``long'' time behaviours. It is worth mentioning here that some known results about the asymptotics of the Mittag-Leffler function relate it to the stretched exponential \cite{mainardi:primo}. This is very interesting in light of the fact that, in the papers by Mignan \cite{mignan:primo,mignan:secondo,mignan:terzo}, the stretched exponential function is indeed expected to give a better fit to the data, especially at large times $(t\gg t_0)$. In the present paper, we have not obtained a rigorous and concrete result about this hint, but it will surely be the object of a future in-depth study.

In order to show a practical application of the theoretical results we obtained, we then explicitly computed the solution of the pure-temporal ETAS rate by using the parameters set values estimated for the Japanese catalog from 1965/01/01 to 2003/09/23 by Zhuang \cite{zhuang:2011}, and pairs of arbitrary fixed values for $(m_0,m_1)$, such to consider the cases $m_0<,=,>m_1$. These order relationships are indeed expected to regulate the possiblity of a break in the decreasing trend of the sequence started by the initial shock, due to the occurrence of ``relevant'' secondary events. We obtained that, if the aftershocks' magnitude $m_1$ is strong, even larger than the initial event's one $m_0$, the behavior of the global temporal rate is supposed to violate the modified Omori law generated by the first event. It consists instead in the superposition of MOLs, which technically leads to a convolution power series. Conversely, if $m_1$ is much smaller than $m_0$, the global temporal rate can be approximated by the modified Omori law started in $t=0$. This follows from the fact that since the aftershocks are very small, they give a negligible contribute to the total temporal rate.

We can conclude that, although the single earthquake sequence we considered here is obviously a plain case for a basic seismic model, still it allowed us to develop a simplified procedure for computing the earthquake rate of a seismic process in a very direct way. More in general, we hope our work will entice to consider fractional calculus to perform theoretical studies of seismic models of epidemic type. As a future work, we aim to generalize the procedure we proposed, to account for randomized aftershocks' magnitudes, that is, consider a generic $m_{\tau}$ in the self-consistent equation~\eqref{SCE}.

\section{Appendix 1: Fractional differential equations} \label{appendix:FDE}

To solve the fractional differential equation in \ref{C}, we refer to Theorem 4.3 in \cite{kilbas:primo}.
We define $C_\gamma[a,b]$ as 
\[C_\gamma[a,b]=\{ f:(a,b] \to \mathbb{R} | (x-a)^\gamma f(x) \in C[a,b]  \}.\] 
In our case, we have $[a,b]=[t_0,T]$ and $(x-t_0)^\gamma f(x)=(x-t_0)^\gamma \frac{1}{(x-t_0)^\theta}=(x-t_0)^{\gamma-\theta}$ which is in $C[t_0,T]$ only if $\gamma -\theta \geq0$, i.e. $\gamma \geq \theta$.
\\
For $\gamma=\theta$ we have that $f(x)\in C_\gamma[a,b]$.

\begin{theorem}\label{KilThm}
	Let $n-1<\alpha<n$ $(n \in \mathbb{N})$ and let $\gamma \in [0,1)$ be such that $\gamma \leq \alpha$.\\
	Also let $\lambda \in \mathbb{R}$. If $f(x) \in C_{\gamma}[a,b]$, the Cauchy problem
	\[ (^cD^\alpha_{a+}y)(x)-\lambda y(x)=f(x), \, x \in [a,b]   \]
	\[ y^{k}(a)=b_k, \, k=0,..,n-1 \]
	with $b_0,..b_{n-1} \in \mathbb{R}$, has a unique solution $y(x)$ which is 
	\[ y(x)=\sum_{j=0}^{n-1}b_j(x-a)^jE_{\alpha,j+1}(\lambda(x-a)^\alpha) + \int_a^x (x-t)^{\alpha -1} E_{\alpha,\alpha}[\lambda(x-t)^\alpha]f(t)dt. \]\\
	where $ E_{\alpha,\alpha}$ is the two-parameter Mittag-Leffler function.\\
\end{theorem}

\section*{Acknowledgements}
The authors would like to thank the Isaac Newton Institute for Mathematical Sciences, Cambridge, for support and hospitality during the programme Fractional Differential Equations [FDE2] where work on this paper was undertaken. This programme was supported by EPSRC grant no EP/R014604/1. Enrico Scalas was also partially supported by the Dr. Perry James (Jim) Browne Research Center at the Department of Mathematics, University of
Sussex. Last but not least, we acknowledge useful discussion with Federico Polito who pointed us to equation \eqref{convolutionintegral}.

This paper is devoted to Francesco Mainardi on the occasion of his 80$^\mathrm{th}$ birthday.

\end{document}